# Evaluating the Application of SOLID Principles in Modern AI Framework Architectures

Jonesh Shrestha
Jarvis College of Computing and Digital Media
DePaul University
Chicago Illinois United States
jshresth@depaul.edu

## ABSTRACT

This research evaluates the extent to which modern AI frameworks, specifically TensorFlow and scikit-learn, adhere to the SOLID design principles—Single Responsibility, Open/Closed, Liskov Substitution, Interface Segregation, and Dependency Inversion. Analyzing the frameworks' architectural documentation and design philosophies, this research investigates architectural trade-offs when balancing software engineering best practices with AI-specific needs. I examined each framework's documentation, source code, and architectural components to evaluate their adherence to these principles. The results show that both frameworks adopt certain aspects of SOLID design principles but make intentional trade-offs to address performance, scalability, and the experimental nature of AI development. TensorFlow focuses on performance and scalability, sometimes sacrificing strict adherence to principles like Single Responsibility and Interface Segregation. While scikit-learn's design philosophy aligns more closely with SOLID principles through consistent interfaces and composition principles, sticking closer to SOLID guidelines but with occasional deviations for performance optimizations and scalability. This research discovered that applying SOLID principles in AI frameworks depends on context, as performance, scalability, and flexibility often require deviations from traditional software engineering principles. This research contributes to understanding how domain-specific constraints influence architectural decisions in modern AI frameworks and how these frameworks strategically adapted design choices to effectively balance these contradicting requirements.

## 1 Introduction

The AI-centric software systems built today depend heavily on modern AI frameworks. However, such frameworks' architectures are rarely assessed with object-oriented principles in mind. This research investigates the adherence to SOLID design principles—Single Responsibility, Open/Closed, Liskov Substitution, Interface Segregation, and Dependency Inversion—in TensorFlow and scikit-learn architectures [3].

TensorFlow is a large-scale machine learning framework that uses dataflow graphs to construct models. It supports execution on distributed systems and specialized hardware [1]. Conversely, scikit-learn is optimized for medium-scale learning problems and strives for consistency and usability through the Estimators for training and predictions [2]. Both frameworks represent different perspectives of AI system design with open-source and well-documented codebases and design principles, which makes them suitable for architectural evaluation.

The increasing popularity of AI has led to a greater dependence on these frameworks across the industry. The SOLID principles guide developers in creating robust, adaptable systems with empirical assessments showing improved software quality, scalability, and code understanding [4,5]. However, AI frameworks must also meet the unique experimental demands of modern AI systems. This research investigates two main questions: (1) to what extent do leading AI frameworks like TensorFlow and scikit-learn adhere to SOLID principles, and (2) what architectural trade-offs are made when balancing software engineering principles with AI-specific requirements. Understanding the balance these frameworks strike between rigorous adherence to the SOLID design principles and practical implications is the objective of this research.

## 2 Literature Review

The SOLID principles comprise five foundational design guidelines to develop maintainable software with object-oriented programming. These include Single Responsibility Principle (SRP), which recommends that every class should have only one reason to change; the Open/Closed Principle (OCP), which allows for class extension without modification; the Liskov Substitution Principle (LSP), which ensures subclasses are substitutable for superclasses; Interface Segregation Principle (ISP), where the class should not be forced to depend on the interfaces they do not use; and Dependency Inversion Principle (DIP), which emphasizes that the dependency should be on abstraction [3]. AI-centric software systems built today depend heavily on modern AI frameworks. However, such frameworks' architectures are rarely assessed with object-oriented principles in mind. As AI systems continue to scale in complexity and size, both vertical and horizontal scaling techniques in AI architectures benefit from modular design patterns that align with the SOLID principles [10].

Empirical studies examining the effects of SOLID design principles found concrete evidence that these principles greatly improve software reusability, maintainability, and scalability. The application of these design principles significantly improved the comprehensibility of machine learning code, with developers spending comparatively less time understanding code that followed SOLID compared to code that did not adhere to these principles [4,5].

Modern AI frameworks are built to support multiple applications focusing on high performance and scalability. TensorFlow uses a dataflow programming model with stateful computation graphs, enabling efficient distributed computing but introducing programming challenges [1]. Alternatively, scikit-learn provides a more standard API for easier integration and use with the rest of the scientific computing environment in Python, like NumPy and SciPy [2]. Its design principles include consistency, inspection, non-proliferation of classes, composition, and sensible defaults, which can be mapped to SOLID principles [7]. Recently, there has been a noticeable shift to more modular design and acceptance of architectural design patterns. Regardless of the unique constraints of AI systems, there has been a wider recognition of software engineering principles in AI development [6]. This shows the importance of research specifically focusing on SOLID principles application within leading AI frameworks.

## 3 Methodology

This research evaluates the adherence to SOLID principles in modern AI framework architectures through a detailed analysis of TensorFlow and scikit-learn. These frameworks were selected based on their popularity, mature open-source codebases, and contrasting architectural approaches—TensorFlow uses a dataflow graph model, and scikit-learn follows a more traditional object-oriented design.

The methodology consists of three primary phases:

1. Analysis of official documentation, including architecture documentation and design principles, academic papers, and online resources such as their GitHub code repositories to understand design philosophies and architectural patterns.
2. Evaluation of source code organization, module structures, architectural components, and their interactions within each framework. For TensorFlow, this involves analyzing Graph, Tensor, and `keras.Model` classes, while in scikit-learn, the focus is on key interfaces like `BaseEstimator` and estimator class hierarchies, which are foundational to each framework.
3. Assessment of framework architecture against each SOLID principle:
   - Single Responsibility: Examining class and module cohesion.
   - Open/Closed: Analyzing extension mechanisms and customization flexibility.
   - Liskov Substitution: Evaluating behavior consistency across inheritance hierarchies.
   - Interface Segregation: Assessing interface separation and abstraction mechanisms.
   - Dependency Inversion: Investigating dependency management approaches.

Throughout all phases, trade-off analysis is conducted where SOLID principles are potentially compromised yet justified for scalability and performance optimization, acknowledging the unique requirements of AI systems.

## 4 Adherence of AI Frameworks to SOLID

This section examines the compliance of each SOLID principle for TensorFlow and scikit-learn framework architecture.

### 4.1 Single Responsibility Principle (SRP)

In TensorFlow, classes have a clear separation of responsibilities as illustrated in Figure 1. `tf.Tensor` manages data flow between nodes, `tf.Session` handles resource distribution across computing hardware like CPUs, GPUs, and TPUs without interfering with graph logic and `tf.Graph` represents mathematical computations where each node performs one specific operation, such as matrix multiplication, convolution, or activation functions [8].

Similarly, scikit-learn's "non-proliferation of classes" design philosophy ensures that only learning algorithms use custom classes [7]. Datasets are represented as NumPy arrays or SciPy sparse matrices; transformers preprocess data with the `transform()` method, while estimators learn from data using the `fit()` method, each maintaining a single, well-defined responsibility [9]. For example, Figure 1 shows scikit-learn's Pipeline class, where data, transformers, and estimators have a single responsibility but can still be chained together to streamline machine learning workflows.

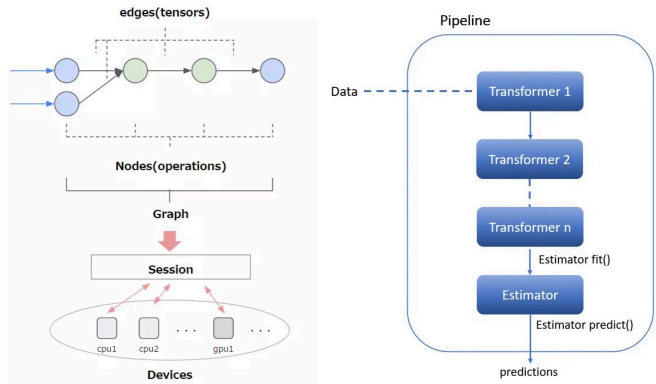

**Figure 1: Comparative Architectural Designs of TensorFlow and scikit-learn [11, 12].**

## 4.2 Open/Closed Principle (OCP)

In TensorFlow Keras API, developers can extend functionality by subclassing `tf.keras.layers.Layer` without modification to create custom layers. This maintains compatibility with existing model structures and seamlessly integrates GPU acceleration [8].

Similarly, scikit-learn allows new estimators to be created by subclassing base classes like `BaseEstimator`. Also, the framework's composition principle facilitates building algorithm sequences using its Pipeline class to chain transformers and estimators [7]. However, scikit-learn lacks native GPU support, operating primarily on the CPU [9]. This hardware resource constraint makes extending scikit-learn for different computing resources challenging, potentially compromising its adherence to the Open/Closed Principle.

## 4.3 Liskov Substitution Principle (LSP)

TensorFlow follows a unified inheritance hierarchy where all layer types like Dense, Conv2D, and LSTM inherit from the base `tf.keras.layers.Layer` class and maintain consistent APIs [8]. This design ensures that different layer types can be substituted for one another in a model, as shown in the code below, where dense and convolution layers follow the same interface despite being different models. This does not violate the computational graph's integrity, provided dimensionality constraints are respected.

```
dense_layer = tf.keras.layers.Dense(64, activation='relu')
conv_layer = tf.keras.layers.Conv2D(64, (3, 3), activation='relu')
input_tensor = tf.random.normal([1, 28, 28, 1])
output1 = dense_layer(tf.reshape(input_tensor, [1, 784]))
output2 = conv_layer(input_tensor)
```

Scikit-learn's `BaseEstimator` implements standard method signatures such as `fit()` and `predict()` across all estimator implementations. Similar to TensorFlow, all models in scikit-learn like `LinearRegression` and `DecisionTreeClassifier` inherit from this common base class and implement `fit(X, y)` and `predict(X)`, ensuring they can be substituted for each other [9]. For instance, in the code below, this uniform interface allows any estimator to be used interchangeably within pipelines, regardless of the underlying model's implementation.

```
X, y = make_classification(n_samples=100, n_features=2, random_state=42)
X_train, X_test, y_train, y_test = train_test_split(
    X, y, test_size=0.3, random_state=42
)
def train_and_predict(model):
    model.fit(X_train, y_train)
    return model.predict(X_test)
train_and_predict(LogisticRegression())
train_and_predict(DecisionTreeClassifier())
```

## 4.4 Interface Segregation Principle (ISP)

TensorFlow implements ISP through domain-specific APIs where `tf.data` is specialized for data loading and preprocessing, `tf.keras` for model building, and `tf.lite` is optimized for mobile deployment where unrelated dependencies like heavy computational graph tools meant for model training are not included. This separation prevents "interface pollution" by ensuring developers only include dependencies relevant to their needs [8]. However, simple layers don't require all the methods inherited from `tf.Modules`, which suggest potential violations.

As shown in the code below, scikit-learn follows consistent API patterns where all estimators use `fit()` for training, transformers use `transform()`, classifiers and regressors use `predict()`, and only probability-supporting classifiers implement `predict_proba()` method [9].

```
scaler = StandardScaler().fit(X_train)
X_scaled = scaler.transform(X_test)
model = RandomForestClassifier().fit(X_train, y_train)
predictions = model.predict(X_test)
probabilities = model.predict_proba(X_test)
```

However, transformers like `PCA` and `StandardScaler` inherit `predict()` despite not using it, and some classifiers must implement the irrelevant `predict_proba()` method because of the `ClassifierMixin` class. This forces clients to depend on methods they don't use.

## 4.5 Dependency Inversion Principle (DIP)

TensorFlow's high-level modules depend on abstractions rather than concrete implementations. For example, high-level API `tf.keras` depends on the `tf.Module` abstraction rather than specific device implementations [1]. The following code snippet demonstrates how TensorFlow provides hardware abstraction through `tf.device`, allowing the same code to run across CPU, GPU, or TPU. Callbacks such as `ModelCheckpoint` and `EarlyStopping` depend on abstract training events like `val_loss`, not specific implementations, making them usable with any model architecture [8].

```
with tf.device('/GPU:0'):
    model = tf.keras.Sequential([
        tf.keras.layers.Dense(128, activation='relu'),
        tf.keras.layers.Dense(10, activation='softmax')
    ])
early_stopping = tf.keras.callbacks.EarlyStopping(monitor='val_loss', patience=3)
model.fit(
    X_train, y_train, validation_data=(X_val, y_val),
    epochs=10, callbacks=[early_stopping]
)
```

As shown in the following code, scikit-learn's preprocessing steps, like `StandardScaler` and `PCA`, depend on an abstract transformer interface like fit and transform, not specific models [9]. Similarly, pipelines rely on the `BaseEstimator` interface, not specific estimator classes, because they only need the standard methods (fit and predict/transform) [7].

```python
X_train = StandardScaler().fit_transform(X_train)
X_train = PCA(n_components=2).fit_transform(X_train)
for model in [LogisticRegression(), RandomForestClassifier()]:
    model.fit(X_train, y_train)
pipeline = Pipeline([
    ('scaler', StandardScaler()),
    ('pca', PCA(n_components=10)),
    ('classifier', LogisticRegression())
])
```

## 5 Trade-offs in AI Framework Architectures

This section discusses how modern AI frameworks balance SOLID principles against performance and scalability demands in AI systems.

### 5.1 Performance versus Modularity

AI frameworks often require balancing strict SOLID design principles with performance requirements. TensorFlow's computational graphs enable efficient execution on GPUs and TPUs but lead to tighter coupling and complexity, violating the SRP [10]. For instance, TensorFlow's Dense layers handle multiple responsibilities (weights management, activation computation, regularization), and Adam optimizer manages (learning rates, gradient updates, optimization state) to improve performance rather than maintain strict modularity [8].

Similarly, scikit-learn's dependence on NumPy arrays shows a practical trade-off for performance and simplicity [2]. While this is a specific implementation rather than an abstraction that violates DIP, it significantly improves performance.

### 5.2 Flexibility versus Computational Efficiency

Balancing flexibility and computational efficiency in AI framework is often important. TensorFlow's graph execution mode optimizes performance but costs flexibility. However, the eager execution introduced in TensorFlow 2.0 improved flexibility at the cost of some performance overhead. The `@tf.function` decorator demonstrates this trade-off, which allows developers to write code in an eager style but compile and execute it as a graph [8]. The `@tf.function` converts Python functions to graph mode, which causes our code to become dependent on TensorFlow's graph execution behavior that relies on low-level implementation [8,10]. This delivers performance benefits but violates some aspects of OCP and DIP.

As shown in the code below, TensorFlow also allows the creation of custom operations that adhere to OCP. However, for performance-critical applications, this approach may not be optimal. In high-performance scenarios, developers might need to implement custom C++ operations or modify TensorFlow's core code [8].

```python
@tf.function
def custom_activation(x):
    return tf.where(x > 0, x, 0.01 * x)

model = tf.keras.Sequential([tf.keras.layers.Dense(128),
    tf.keras.layers.Lambda(custom_activation),
    tf.keras.layers.Dense(10, activation='softmax')
])
```

### 5.3 Maintainability versus Scalability

The experimental nature of AI development creates unique challenges in maintaining a clean framework architecture that adheres to SOLID principles while also enabling scalability. TensorFlow's distributed execution capabilities across diverse hardware provide exceptional scalability but introduce additional complexity and coupling between components that reduce maintainability [10].

Similarly, Tensorflow and scikit-learn's uniform interfaces sometimes include methods that aren't always used. The framework avoids the overhead of managing several smaller interfaces, which reduces the complexity of method calls, which slightly violates ISP but provides consistency that improves overall maintainability [8,9].

## 6 Conclusion

TensorFlow prioritizes performance and scalability for deep learning applications, often at the expense of strict SOLID adherence. Nonetheless, TensorFlow's strong performance in real-world applications shows that in performance-critical domains, such trade-offs are justifiable. Meanwhile, the high-level Keras API offers a more maintainable option at the cost of performance that aligns better with SOLID principles [8].

Alternatively, scikit-learn is designed with principles that align more closely with SOLID, like consistency, inspection, and composition [7]. Its architecture prioritizes consistent, simple interfaces for maintainability and ease of use, making it suitable for classical ML projects. However, it may require additional optimizations for very large datasets, potentially compromising some design principles for performance [9].

In conclusion, AI development is iterative and experimental, with the constant need to tweak data, models, and algorithms, unlike traditional software development, which assumes a stable problem. This makes applying SOLID principles tricky in AI development [10]. To adapt, libraries like TensorFlow and scikit-learn focus on flexible designs, making architectural decisions that prioritize scaling to large datasets and complex models, sometimes at the expense of perfect modularity or substitutability. For instance, the distributed computing capabilities found in AI frameworks boost performance across multiple machines but add complexity through additional coupling between components. Hence, the application of SOLID principles in AI frameworks requires a thoughtful consideration of domain-specific needs rather than strict adherence to SOLID principles.